\documentclass{emulateapj}

\shorttitle{Obscured AGNs in the CDFS}
\shortauthors{Fiore et al.}

\def\ls{{_<\atop^{\sim}}}
\def\gs{{_>\atop^{\sim}}}
\def\cgs{ ${\rm erg~cm}^{-2}~{\rm s}^{-1}$ }

\begin{document}

\title{Unveiling obscured accretion in the Chandra Deep Field South}
\author{F. Fiore\altaffilmark{1}, A. Grazian\altaffilmark{1},
P. Santini\altaffilmark{1,2}, S. Puccetti\altaffilmark{3},
M. Brusa\altaffilmark{4}, C. Feruglio\altaffilmark{1},
A. Fontana\altaffilmark{1}, E. Giallongo\altaffilmark{1},
A. Comastri\altaffilmark{5}, C. Gruppioni\altaffilmark{5}, 
F. Pozzi\altaffilmark{6}, G. Zamorani\altaffilmark{5}, 
C. Vignali\altaffilmark{6}}
\altaffiltext{1}{INAF-OAR, via Frascati 33, Monteporzio, I00040, Italy}
\altaffiltext{2}{Universita' di Roma La Sapienza, Italy} 
\altaffiltext{3}{ASI Science data Center, via Galileo Galilei, 
00044 Frascati, Italy}
\altaffiltext{4}{Max Planck Institut f\"ur extraterrestrische Physik,
      Giessenbachstrasse 1, D--85748 Garching bei M\"unchen, Germany}
\altaffiltext{5}{INAF-OABo, via Ranzani 1, Bologna, Italy}
\altaffiltext{6}{Universita' di Bologna, via Ranzani 1, Bologna, Italy}

\email{fiore@oa-roma.inaf.it}

\begin{abstract}
We make use of deep HST, VLT, Spitzer and Chandra data on the Chandra
Deep Field South to constrain the number of Compton thick AGN in this
field.  We show that sources with high 24$\mu$m to optical flux ratios
and red colors form a distinct source population, and that their
infrared luminosity is dominated by AGN emission. Analysis of the
X-ray properties of these extreme sources shows that most of them
(80$\pm15\%$) are indeed likely to be highly obscured, Compton thick
AGNs. The number of infrared selected, Compton thick AGNs with
5.8$\mu$m luminosity higher than $10^{44.2}$ erg s$^{-1}$ turns out to
be similar to that of X-ray selected, unobscured and
moderately obscured AGNs with 2-10 keV luminosity higher than
$10^{43}$ erg s$^{-1}$ in the redshift bin 1.2-2.6. This ``factor of
2'' source population is exactly what it is needed to solve the
discrepancies between model predictions and X-ray AGN selection.

\end{abstract}



\section{Introduction}

Active Galactic Nuclei (AGN) are not only witnesses of the phases of
galaxy formation and/or assembly, but are most likely among leading
actors. Indeed, three seminal discoveries indicate tight links and
feedbacks between super-massive black holes (SMBH), nuclear activity
and galaxy evolution. The first is the discovery of SMBH in the center
of most nearby bulge dominated galaxies, and the tight correlation
between their masses and galaxy bulge properties (Gebhardt et
al. 2000, Ferrarese \& Merritt 2000, Marconi \& Hunt 2003 and
references therein).  The second is that the growth of SMBH is mostly
due to accretion of matter during their active phases, and therefore
that most bulge galaxies passed a phase of strong nuclear activity
(Soltan 1982, Marconi et al. 2004).  The third is that the evolution
of AGN is luminosity dependent, with lower luminosity AGN peaking at a
redshift lower than luminous QSOs (Hasinger 2003, 2005, Fiore et
al. 2003, Ueda et al. 2003, La Franca et al. 2005, Brandt \& Hasinger
2005, Bongiorno et al. 2007), a bimodal behavior recalling the
evolution of star-forming galaxies and that of massive spheroids
(Cowie et al. 1996, Franceschini et al. 1999, De Lucia et al. 2006).
All three discoveries imply that obtaining a complete census of
accreting SMBH through the cosmic epochs and constraining accretion
efficiency and feedbacks are crucial steps toward the understanding of
Galaxy formation and evolution.

First attempts to constrain models for the formation and evolution of
structure in the Universe using the evolving optical and X-ray AGN
luminosity functions have been presented by Granato et al. (2001,
2004), Di Matteo et al. (2005), and Menci et al. (2004, 2005). In
particular, the Menci et al. model links the evolution of the galaxies
in the hierarchical clustering scenarios with the changing accretion
rates of cold gas onto the central SMBH that powers the QSO (Cavaliere
\& Vittorini 2000).
The results of this model were encouraging, in the sense that it
predicts a trend of lower luminosity AGN to peak at increasingly lower
redshift, as observed. However, from a quantitative point of view, the
model over-predicts by a factor of about 2 the space density of
low-to-intermediate luminosity (Seyfert like) AGNs at z=1.5-2.5 with
respect to present X-ray observations.  Furthermore, Marconi
et. al. (2004, 2007 in preparation) derived a SMBH mass function from
the X-ray selected AGN luminosity functions (e.g.  La Franca et
al. 2005) that falls short by a factor of about 2 to the ``relic''
SMBH mass function, evaluated using the M$_{BH} - \sigma_V$ / M$_{BH}
-$M$_B$ relationships and the local bulge's luminosity function.  The
most likely explanation for both discrepancies is that present X-ray
surveys, although very efficient to probe unobscured and moderately
obscured AGN (with column densities up to a few $10^{23}$ cm$^{-2}$,
the so-called Compton thin AGNs), miss most of the very highly
obscured, but still strongly accreting objects, the so called Compton
Thick AGNs, with a column density N$_H\gs10^{24}$ cm$^{-2}$ (see
Comastri 2004). Indeed, only a handful of the faintest sources in the
Chandra deep fields may be Compton thick (see La Franca et al. 2005
and Tozzi et al. 2006).  So we still may be viewing just the tip of
the iceberg of the highly obscured AGN population. Compton thick
objects may well be more common at high redshift, as suggested on
theoretical ground by Silk \& Rees (1998) and Fabian (1999) and on
observational ground by e.g. Gilli et al. (2001), Worsley et
al. (2004, 2006), and La Franca et al. (2005).

Compton thick AGN at z$\gs$1 can be recovered thanks to the
reprocessing of the AGN UV emission in the infrared by selecting
sources with AGN luminosities in the mid-infrared and faint
near-infrared and optical emission (e.g. Martinez-Sansigre et
al. 2005, 2006, Houck et al. 2005, Weedman et al. 2006a, 2006b).  We
investigate further this issue making use of the multiwavelength data
obtained on the Chandra Deep Field South (Giacconi et al. 2002), one
of the fields with the deepest coverage at optical, infrared and X-ray
wavelengths.  

A $H_0=70$ km s$^{-1}$ Mpc$^{-1}$, $\Omega_M$=0.3,
$\Omega_{\Lambda}=0.7$ cosmology is adopted throughout.

\section{Datasets and sample selection}

\begin{figure}
\includegraphics[height=8.5truecm,width=8.5truecm]{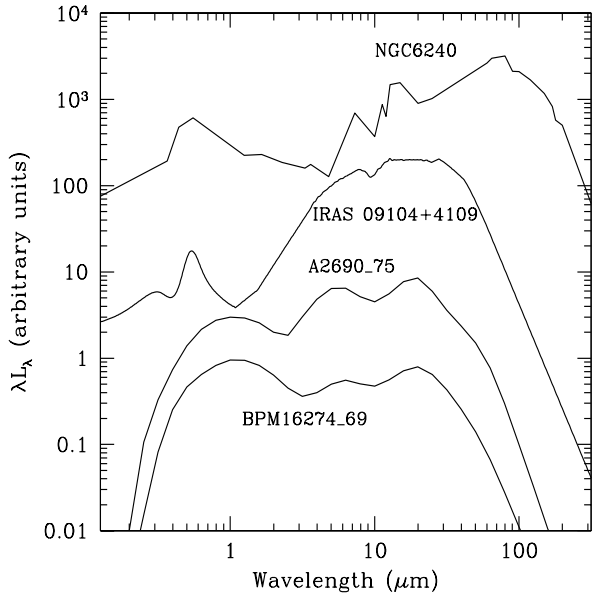}
\caption{$\lambda L_\lambda$ spectral energy distributions of the four 
additional templates of sources hosting highly obscured AGNs used in this work.
From top to bottom: NGC6240; IRAS09104+41091; 
HELLAS2XMM A2690\_75 and BPM16274\_69}
\label{seds}
\end{figure}

The selection and spectroscopic identification of complete AGN samples
from mid-IR surveys is a rather difficult task, because AGNs make only
a small fraction of the full mid-IR source population. Using Spitzer
IRAC colors results in samples significantly contaminated by
star-forming galaxies (e.g. Lacy et al. 2004, Alonso-Herrero et
al. 2006, Barmby et al. 2006, Polletta et al. 2006). Comparing the
global observed Spectral Energy Distribution (SED) to AGN and galaxy
templates proved to be more efficient in selecting AGN samples
(Polletta et al. 2006, 2007). However, it is very difficult to assess
the completeness of these samples given all complex selection
effects. Furthermore, many obscured AGN may still be anyway missed by
this technique. 

We adopt in this paper a somewhat different approach. We do not
pretend to select {\it all} AGN through optical and infrared
photometry. As explained before X-rays are much more efficient in
selecting unobscured (i.e. broad line AGNs) and moderately obscured
AGNs. We concentrate our effort on highly obscured AGN only and we
limit our analysis to the high infrared luminosity AGN population. The
driving consideration is that differences between nuclear and
star-formation emission are emphasized comparing the observed SEDs
with galaxy templates over a range as broad as possible (also see
Martinez-Sansigre et al. 2005,2006, Houck et al. 2005 and Weedman et
al. 2006, Magliocchetti et al. 2007). Our primary objective here
is to validate our obscured AGN selection criteria and assess the
magnitude of the corresponding selection effects using a careful
analysis of the deep X-ray data available on the CDFS.

\begin{figure*}[t]
\begin{tabular}{cc}
\includegraphics[height=8.5truecm,width=8.5truecm]{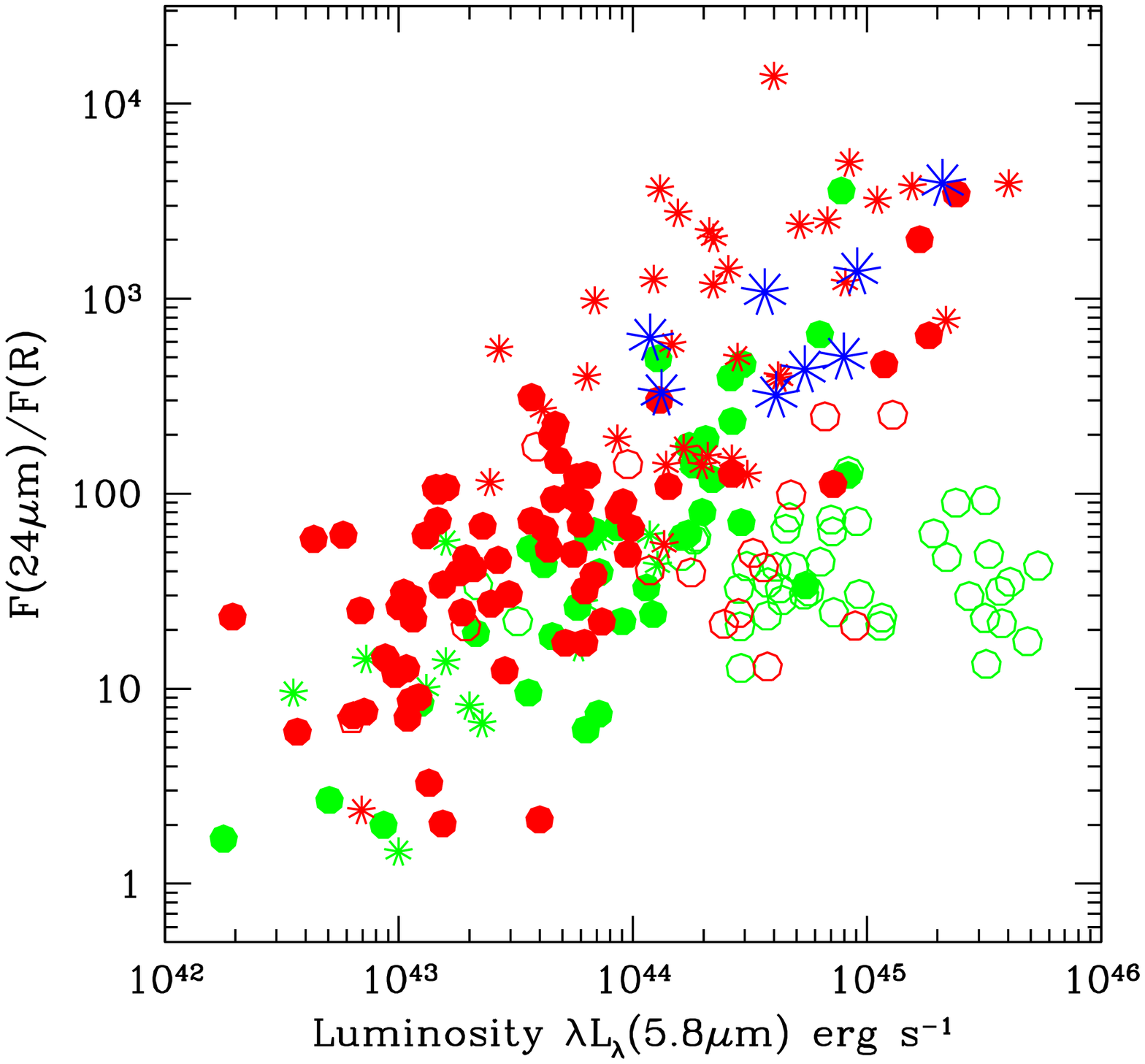}
\includegraphics[height=8.5truecm,width=8.5truecm]{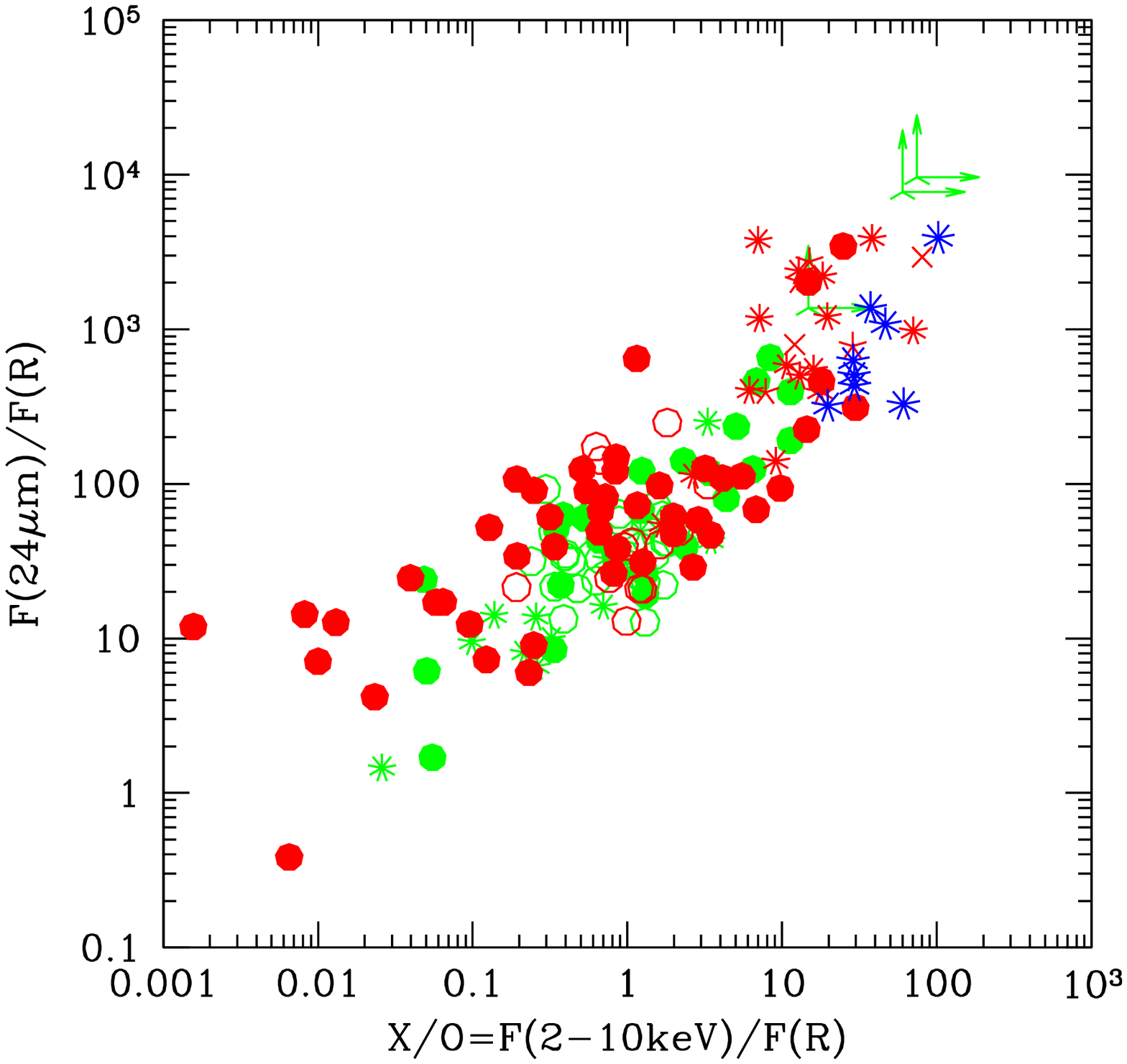}
\end{tabular}
\caption{Left panel: F($24\mu$m)/F(R) as a function of the 5.8$\mu$m
luminosity for three X-ray source samples (GOODS-MUSIC, red symbols,
ELAIS-S1, green symbols and HELLAS2XMM, large blue symbols , Pozzi et
al. 2007).  Open circles = type 1 AGN; filled circles = non type 1
AGN; stars = photometric redshifts.  Note that F($24\mu$m)/F(R) of non
broad line AGN is strongly correlated with the luminosity at
5.8$\mu$m. Right panel: F($24\mu$m)/F(R) as a function of the X-ray to
optical flux ratio X/O for the same source samples. All symbols as in left
panel; skeleton triangles = sources without optical counterpart.}
\label{mirolir}
\end{figure*}

\subsection{The GOODS-MUSIC catalog}

We use in this paper the latest version of the GOODS-MUSIC catalog
(Grazian et al. 2006). We limit the analysis to the region fully
covered by deep VLT/ISAAC near infrared photometry (143.2 arcmin$^2$)
and to the sources with MIPS 24$\mu$m fluxes F(24$\mu$m)$>40\mu$Jy
(1729 sources). 24$\mu$m fluxes for all objects
in the catalog, were obtained following the procedures described in De
Santis et al. (2007).  

The published version of the catalog contains z band and K band
selected sources in the GOODS-south area. The revised version of the
catalog used in this work includes 46 objects that are detected only
in the 4.5$\mu$m band, i.e. their z and K magnitudes are below the
chosen detection threshold. Only 4 of these sources do not have
counterparts in the optical and/or in the near infrared images.  At
the flux limits adopted here, we do not detect any objects at 24$\mu$m
that is not detected at shorter wavelengths. In the following we use for each
entry of the catalog Vega magnitudes and cgs fluxes.  

The GOODS-MUSIC catalog includes both monochromatic and total
infrared luminosities, for the sources with a reliable spectroscopic
or photometric redshifts.  Total 8--1000$\mu$m luminosities were
computed by integrating the best fit galaxy and AGN templates.
Monochromatic luminosities were computed by interpolating the observed
SEDs at the rest frame wavelength of interest.  A large (factor of
10-30) systematic uncertainty is associated to the total infrared
luminosity, being this dominated by the contributions at wavelengths
of $\sim100-1000\mu$m, well outside the infrared band used in this
paper (1-24 $\mu$m).  Different models can produce similar fits
below 24$\mu$m but give rise to large differences in the total
infrared luminosity. This systematic uncertainty is greatly reduced
using the monochromatic luminosity at 5.8 $\mu$m (see e.g. Yan et
al. 2007), since this wavelength is within the observed band up to
z$\sim$3.2. For this reason in the following we make use of the
infrared luminosity at 5.8$\mu$m ($\lambda L_\lambda (5.8\mu$m)) to
characterize the infrared power of the CDFS sources.

\subsection{Galaxy and AGN templates}

A detailed fitting of the observed SEDs using AGN and galaxy templates
was also performed.  Templates include passive galaxies, star-forming
galaxies, unobscured AGNs and highly obscured AGNs. Spectral
libraries include both empirical templates (Coleman et al. 1980 datasets,
Polletta et al. 2007) and synthetic models (Bruzual \& Charlot 2003,
Fioc \& Rocca-Volmerange 1997).  We put particular care on the
description of highly obscured AGNs.  In addition to the templates
presented by Polletta et al. (2007) we used four additional templates
of sources hosting highly obscured AGNs (figure \ref{seds}). The NGC6240
template includes U, B, V, J, H, K, IRAS and ISO
photometry. The IRAS09104+41091 includes SDSS u, g, r, i, z
photometry, B, V, R, J, H, K photometry, Spitzer IRAC and MIPS
photometry, Spitzer IRS spectroscopy and IRAS photometry. For the two
HELLAS2XMM sources we use the best fit SED model in Pozzi et
al. (2007). These templates span a range of infrared to optical flux
ratio significantly broader than the obscured AGN templates used by
Polletta el al. (2007).

\subsection {Photometric redshifts and SED fittings}

Photometric redshifts were derived by Grazian et al. (2006) fitting
only the part of the SED dominated by the integrated stellar
population, i.e. $\lambda<5.5 \mu$m (Grazian et al. 2006 and
references therein). The synthetic models used are very accurate and
complete in the treatment of the star-formation history and evolution
of the stellar populations and of the dust content of the galaxy and
its evolution. However, while dust extinction is easy to account for
because it is a line of sight effect, dust emission and reprocessing
is much more complicated to model, since it depends largely on the
assumed geometry and covering fraction. Dust emission was therefore
not considered in the models, and accordingly the bands above
$5.5\mu$m rest frame are ignored in these fits (see Grazian et
al. 2006 for further details).  This approach produces robust
photometric redshifts ($\Delta z/(1+z) <0.05$) for passive galaxies,
star-forming galaxies and obscured AGNs, where the nuclear optical and
infrared emission is completely blocked or strongly reduced by dust
and gas along the line of sight. Unobscured AGNs with power law SEDs
are excluded from this analysis (however, most of them have reliable
spectroscopic redshift from Cimatti et al. 2002, Szokoly et al. 2004,
Le Fevre et al. 2004, Vanzella et al.2005,2006, Mignoli et al. 2005).

To characterize each SED we repeated the fit to the observed SED with
a library of empirical templates, fixing the redshift to the
spectroscopic redshift or, if this is not present, to the photometric
redshift obtained as described above. This library includes the 21
templates of passive galaxies, star-forming galaxies and AGNs by
Polletta et al.  (2007) and the four templates of systems hosting a
highly obscured AGN in figure \ref{seds}.  The best fit template and
normalization were used to measure monochromatic and total infrared
luminosities.  

\begin{figure*}[t!]
\begin{tabular}{cc}
\includegraphics[height=8.5truecm,width=8.5truecm]{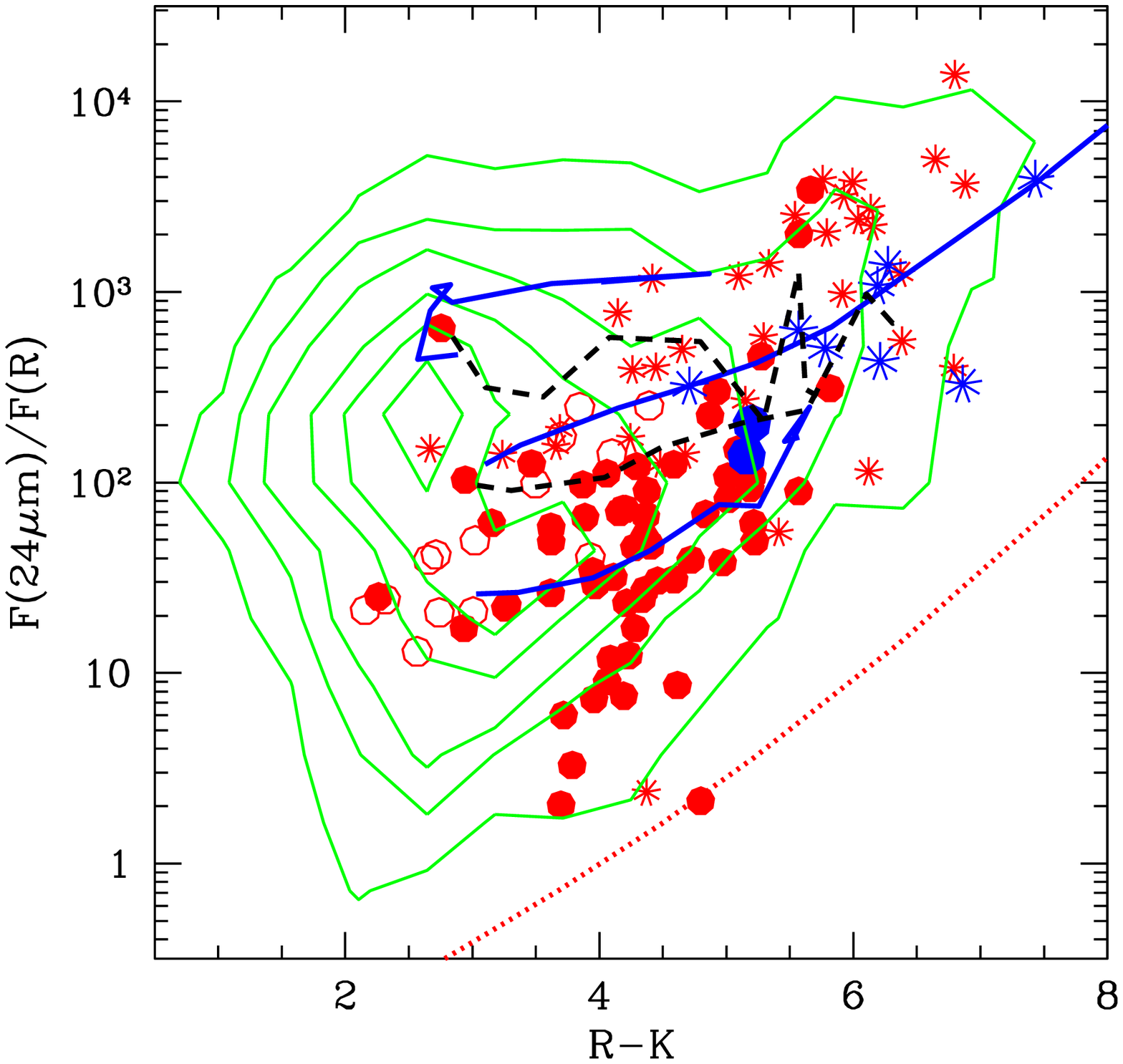}
\includegraphics[height=8.5truecm,width=8.5truecm]{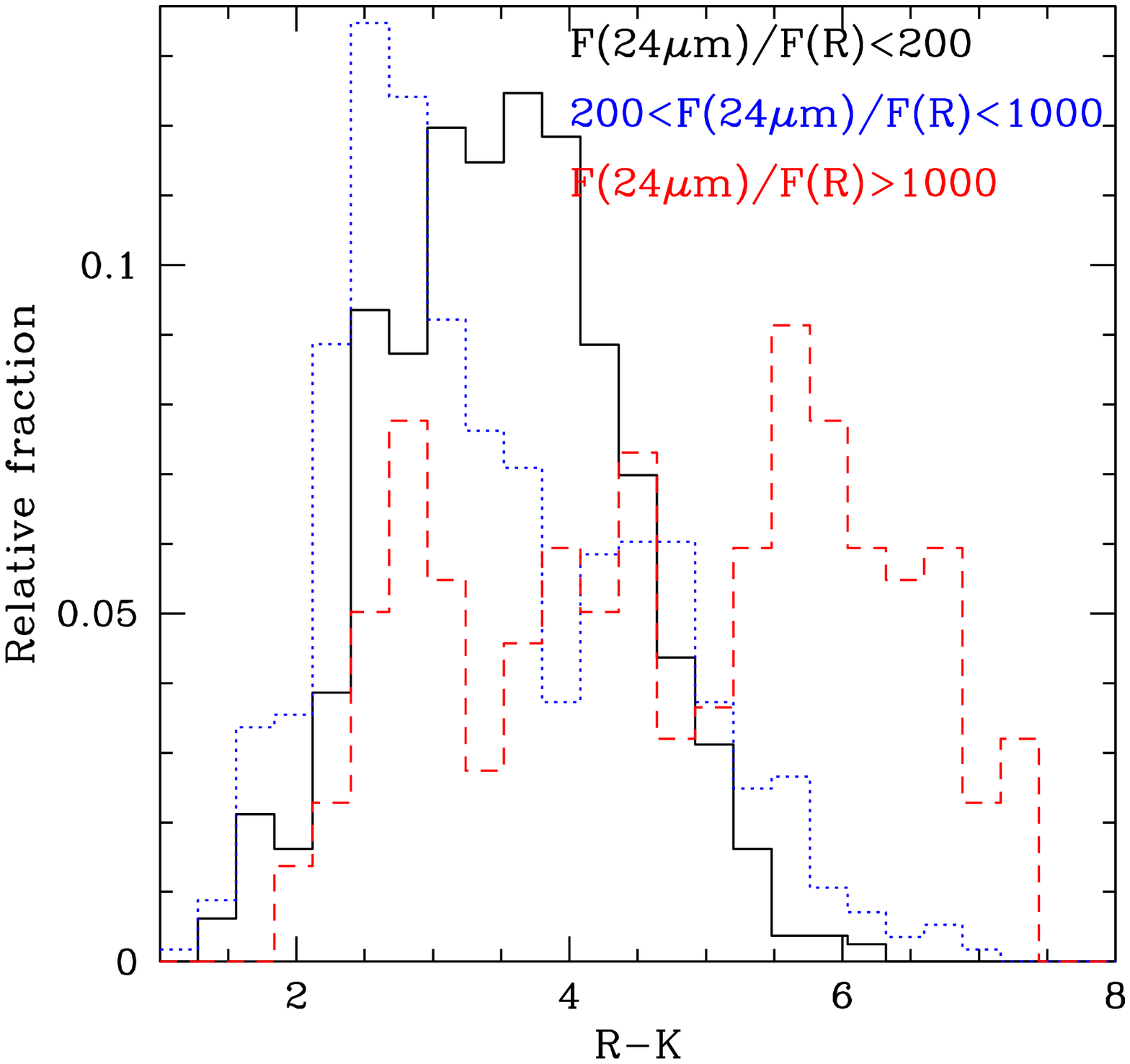}
\end{tabular}
\caption{Left panel: F($24\mu$m)/F(R) as a function of the R-K
color for two X-ray source samples (GOODS-MUSIC and HELLAS2XMM, large
symbols, from Pozzi et al. 2007). Open circles = type 1 AGN; filled
circles = non type 1 AGN; stars = photometric redshifts. Isodensity
contours of all GOODS-MUSIC 24$\mu$m sources are overlayed to the
plot. The thick continuous lines mark the expectations of three
obscured AGN SEDs with redshift increasing from 0 to 4 from left to
right. The lower curve represents the colors of a typical low
luminosity Seyfert 2 galaxy, the middle curve represents the colors of
an obscured AGN from the Pozzi et al sample (A2690\_75), the
upper curve represents the colors of IRAS09104+41091.  The black dashed
lines are the expectations of the SEDs of two star-burst galaxies
(M82, lower curve and Arp220, upper curve) for z=0-4.  The dotted line
is the expectation of a passive elliptical galaxy for z=0-4.  Right
panel: fraction of GOODS-MUSIC 24$\mu$m sources as a function of the
R-K color in three F($24\mu$m)/F(R) bins: solid histogram =
F($24\mu$m)/F(R)$<200$; dotted histogram =
$200<$F($24\mu$m)/F(R)$<1000$; dashed histogram =
F($24\mu$m)/F(R)$>1000$.}
\label{mirormk}
\end{figure*}

\subsection{Optical, near infrared and mid infrared color selection}

The longest and shorthest wavelengths at which deep photometry is
available on the Chandra Deep Fields are the 24$\mu$m mid-infrared
band covered by Spitzer MIPS and the optical bands covered by HST ACS.
24$\mu$m sources with faint optical counterparts must be either
luminous AGN whose optical nuclear emission is blocked by dust and
gas, or powerful dusty-starburst galaxies. The Mid-infrared to optical
flux ratio (\footnote{F(24$\mu$m)/F(R); logF(R)=$-0.4\times
R-22.5467$. R magnitudes have been obtained by interpolating the V and
I magnitudes in the GOODS-MUSIC catalog provided by HST/ACS}) can
therefore be considered a rough estimator of obscured activity (both
nuclear and star-formation) in galaxies.  

Figure \ref{mirolir} shows F(24$\mu$m)/F(R) as a function of $\lambda
L_\lambda (5.8\mu$m), for three samples of X-ray sources.  Unobscured
AGNs (open symbols in figure \ref{mirolir}) have F(24$\mu$m)/F(R) in
the range 10-200, uncorrelated with $\lambda L_\lambda (5.8\mu$m), as
expected because the nuclear emission dominates both optical and mid
infrared wavelengths. Conversely, obscured AGNs (filled symbols) have
F(24$\mu$m)/F(R) spanning a broader range, and fairly correlated with
$\lambda L_\lambda (5.8\mu$m). This behavior resembles that of
moderately obscured AGNs detected in X-rays, for which the X-ray to
optical flux ratio (X/O) is strongly correlated with the X-ray
luminosity (Fiore et al. 2003, Eckart et al. 2006).  The nuclear
optical-UV light of these objects is completely blocked, or strongly
reduced by dust extinction, and the optical flux is dominated by the
host galaxy. On the other hand, the 2-10 keV flux is reduced by only
small factors, even for obscuring gas column densities of the order of
a few$\times 10^{23}$ cm$^{-2}$.  X/O is therefore a good estimator of
the ratio between the nuclear flux and the host galaxy starlight
flux. While the nuclear AGN X-ray luminosity can span several decades,
the host galaxy R band luminosity has a moderate scatter, less than
one decade, giving rise to the observed correlation between X/O and
the X-ray luminosity. Perola et al. (2004), Mignoli et al. (2004),
Brusa et al. (2005) and Cocchia et al. (2007) found that the high
X/O sources tend also to be obscured in the X-rays, with column
densities of the order of $10^{22-23}$ cm$^{-2}$. It is therefore
possible to conclude that a high X/O ratio is a good indicator for
both optical and moderate X-ray obscuration in high luminosity
sources.  Interestingly, the F(24$\mu$m)/F(R) of X-ray selected
sources is strongly correlated with X/O (figure \ref{mirolir}, right
panel). This suggests that luminous Compton thick AGN, which are faint
in X-rays because of the strong photoelectric absorption and Compton
scattering, and cannot be selected using their X/O flux ratio, can be
recovered using the F(24$\mu$m)/F(R) ratio.

\begin{figure*}
\begin{tabular}{cc}
\includegraphics[height=8.5truecm,width=8.5truecm]{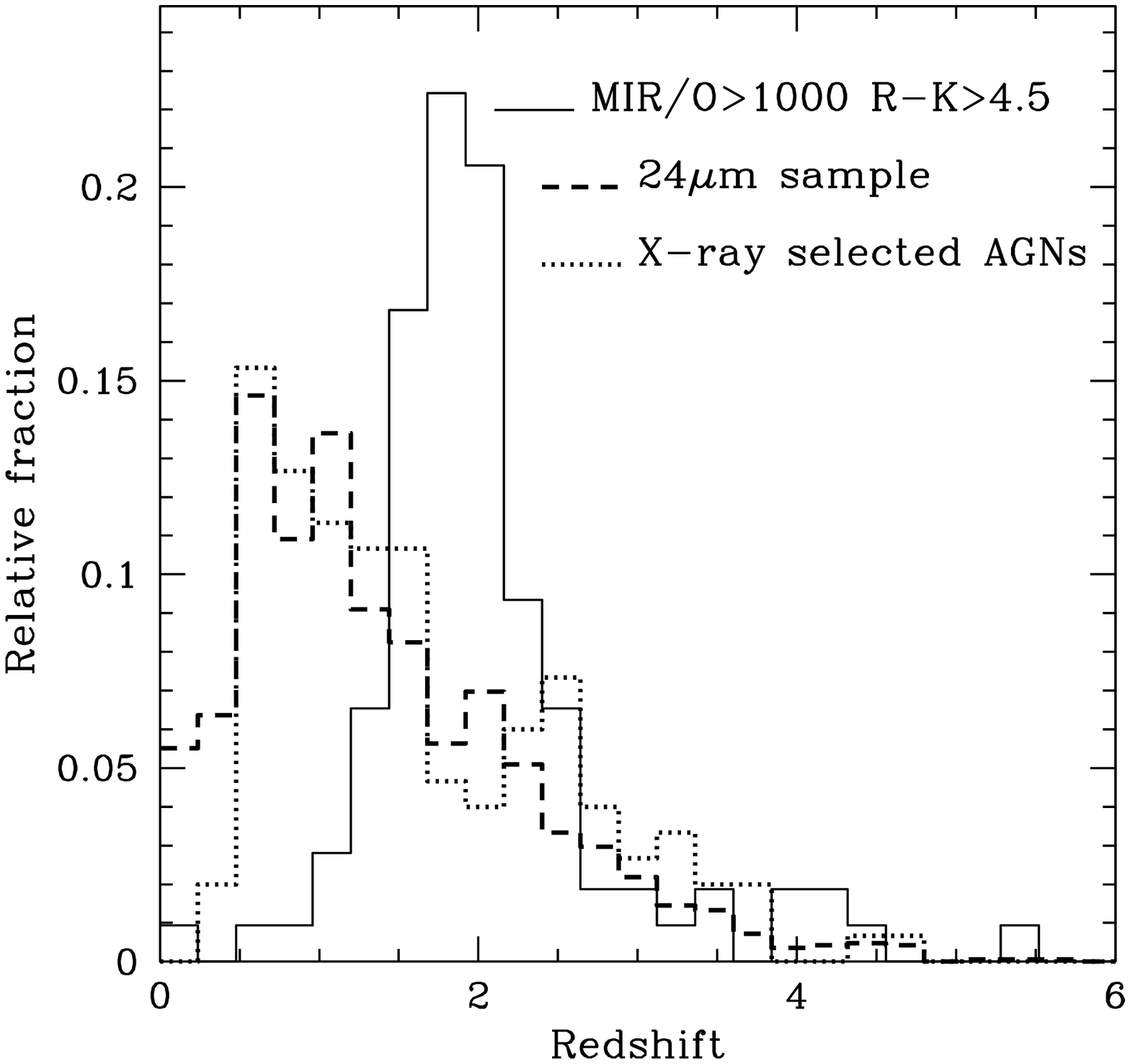}
\includegraphics[height=8.5truecm,width=8.5truecm]{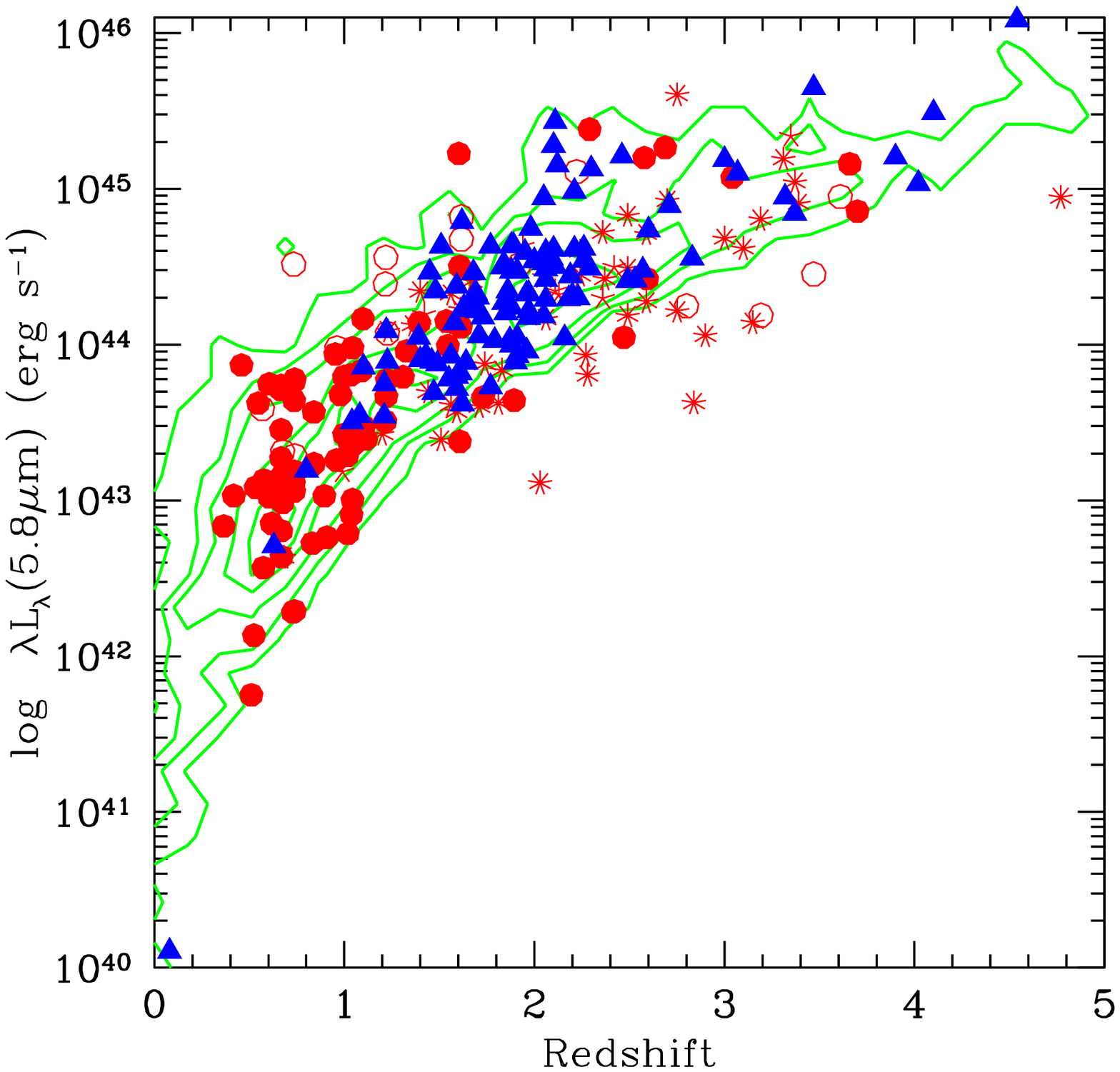}
\end{tabular}
\caption{Left panel: the redshift distribution of the
F(24$\mu$m)/F(R)$>1000$ and R-K$>4.5$ sources without direct X-ray
detection (107 sources) compared to those of the full GOODS-MUSIC
24$\mu$m sample (1649 sources) and of the X-ray selected AGNs in the
GOODS-MUSIC area (150 sources). Right panel: the redshift-infrared
luminosity (log$(\lambda L_\lambda (5.8\mu m))$) plane for the full
24$\mu$m GOODS-MUSIC source sample (isodensity contours); the CDFS
X-ray AGN sample (open circles = type 1 AGN; filled circles = non type
1 AGN; stars = photometric redshifts); and the sample of
F(24$\mu$m)/F(R)$>1000$ and R-K$>4.5$ sources without direct X-ray
detection (filled triangles)}
\label{zdist}
\end{figure*}

Furthermore, since X-ray obscured AGNs tend to have red R-K colors
(Brusa et al. 2005 and reference therein), one would also expect that
Compton thick AGN have similarly red colors.  Indeed figure
\ref{mirormk} shows that the F(24$\mu$m)/F(R) of X-ray selected,
obscured AGN is correlated with the R-K color, as expected.  Figure
\ref{mirormk} also shows the iso-density contours of all the
GOODS-MUSIC 24$\mu$m sources with F(24$\mu$m)$>40\mu$Jy.
Intriguingly, the iso-density contours become narrow in
F(24$\mu$m)/F(R) at high R-K values and extend toward the region
occupied by obscured X-ray selected AGN at high F(24$\mu$m)/F(R) and
high R-K values.  The bimodality at high values of F(24$\mu$m)/F(R) of
the color distribution of the 24$\mu$m selected sources is evident in
the rightmost panel of figure \ref{mirormk}, which shows the fraction
of GOODS-MUSIC 24$\mu$m sources as a function of the R-K color in
three F(24$\mu$m)/F(R) bins. While at low and intermediate
F(24$\mu$m)/F(R) values the distributions are peaked at
R-K$\sim2.5-3.5$ and decrease smoothly toward higher R-K values, the
distribution of the sources with F(24$\mu$m)/F(R)$>1000$ shows a
strong excess at R-K$>4.5$.

It is interesting to note that most of the highly obscured AGN
selected in the HELLAS2XMM survey on the basis of their high X-ray to
optical flux ratio (Pozzi et al. 2007) have F(24$\mu$m)/F(R) higher
than a few hundred, and all have R-K$>4.5$.  Their SEDs are
characterized by a passive early-type galaxy in the optical and in the
near infrared, and by an AGN component in the mid infrared. These SEDs
redshifted up to z=4 are able to explain the extreme colors of the
F(24$\mu$m)/F(R)$>1000$ R-K$>4.5$ sources, unlike the SED of even
extreme star-forming galaxies like Arp220 (see figure \ref{mirormk}
left panel). This strongly suggests that most of the
F(24$\mu$m)/F(R)$>1000$ R-K$>4.5$ sources are powered by an active
nucleus.  Similar conclusions are found analyzing slightly different
color diagrams, like F(24$\mu$m)/F(R) vs. F(24$\mu$m)/F(8$\mu$m) and
F(24$\mu$m)/F(R) vs. F(3.6$\mu$m)/F(z).

\section{X-ray properties of extreme 24$\mu$m selected sources}

Our candidate obscured AGNs are selected from the full 24$\mu$m
GOODS-MUSIC sample with the request of having F(24$\mu$m)/F(R)$>1000$
and R-K$>4.5$. There are 135 such sources.

\subsection{Sources with a direct X-ray detection}

Eighteen of the 135 sources have an X-ray detection in Alexander et
al. (2003).  Four other sources are not formally detected but have
more than 4-5 counts (after background subtraction) at the position of
the 24$\mu$m source. In summary, 22 of the 135 sources (16\%) have a
significant X-ray emission directly visible in the Chandra images.

Three of these 22 sources have a spectroscopic redshift, with narrow
line optical spectra, the other 19 have a photometric
redshift in the GOODS-MUSIC catalog (see Section 2); the median
redshift and its interquartile range of these 22 sources are 2.1 (0.5;
hereinafter interquartile ranges are indicated in brackets after
the median values).

The median monochromatic infrared luminosity at 5.8$\mu$m is 44.42
(0.37).  The X-ray luminosities are in all cases higher than $10^{42}$
erg s$^{-1}$, making them bona-fide AGNs.  The median logarithmic
ratio between the $5.8\mu$m and the 2-10 keV luminosities is 1.07
(0.32). As a comparison, the median of the same logarithmic ratio for
the full GOODS-MUSIC X-ray sample (150 AGN with measured redshift) is
0.69 (0.47). The probability that the two distributions are drawn from
the same parent population is $\ls0.2\%$, using the Kolmogorov-Smirnov
test.

The hardness ratios indicate in most cases a hard, possibly obscured,
X-ray spectrum. Indeed, these sources are among the most obscured ones
in the Tozzi et al (2006) analysis, all having column densities higher
than a few$\times10^{22}$ cm$^{-2}$ and 2 having column densities as
high as $10^{24}$ cm$^{-2}$.  

Table 1 gives the breakdown of the best fitting templates (see Section
2) to the SEDs of these 22 sources. 15 SEDs are best fitted by one of
the templates in figure \ref{seds}.

\begin{table}
\caption{\bf Table 1: template fits to the SEDs of the sources with 
 F(24$\mu$m)/F(R)$>1000$ and R-K$>4.5$}
\begin{tabular}{lcc}
\hline
\hline
Template                    & X-ray det. & Not X-ray det.  \\
\hline
Ellipticals + S0            & - & 2   \\
Spirals                     & - & 1   \\
M82+N6090+Arp220            & - & 35  \\
I19254 + Mark231            & 4 & 17    \\
Seyfert 1.8-2+red QSO       & 1 & 5  \\
A2690\_75 + BPM16274\_69    & 5 & 34  \\
IRAS09104+4109              & 9 & 11  \\
N6240                       & 1 & 1   \\
Seyfert 1 +QSOs             & 2 & 1   \\
\hline
Total                       & 22 & 107 \\
\hline
\end{tabular}

\end{table}

\begin{figure*}
\begin{tabular}{cc}
\includegraphics[height=8truecm,width=8truecm]{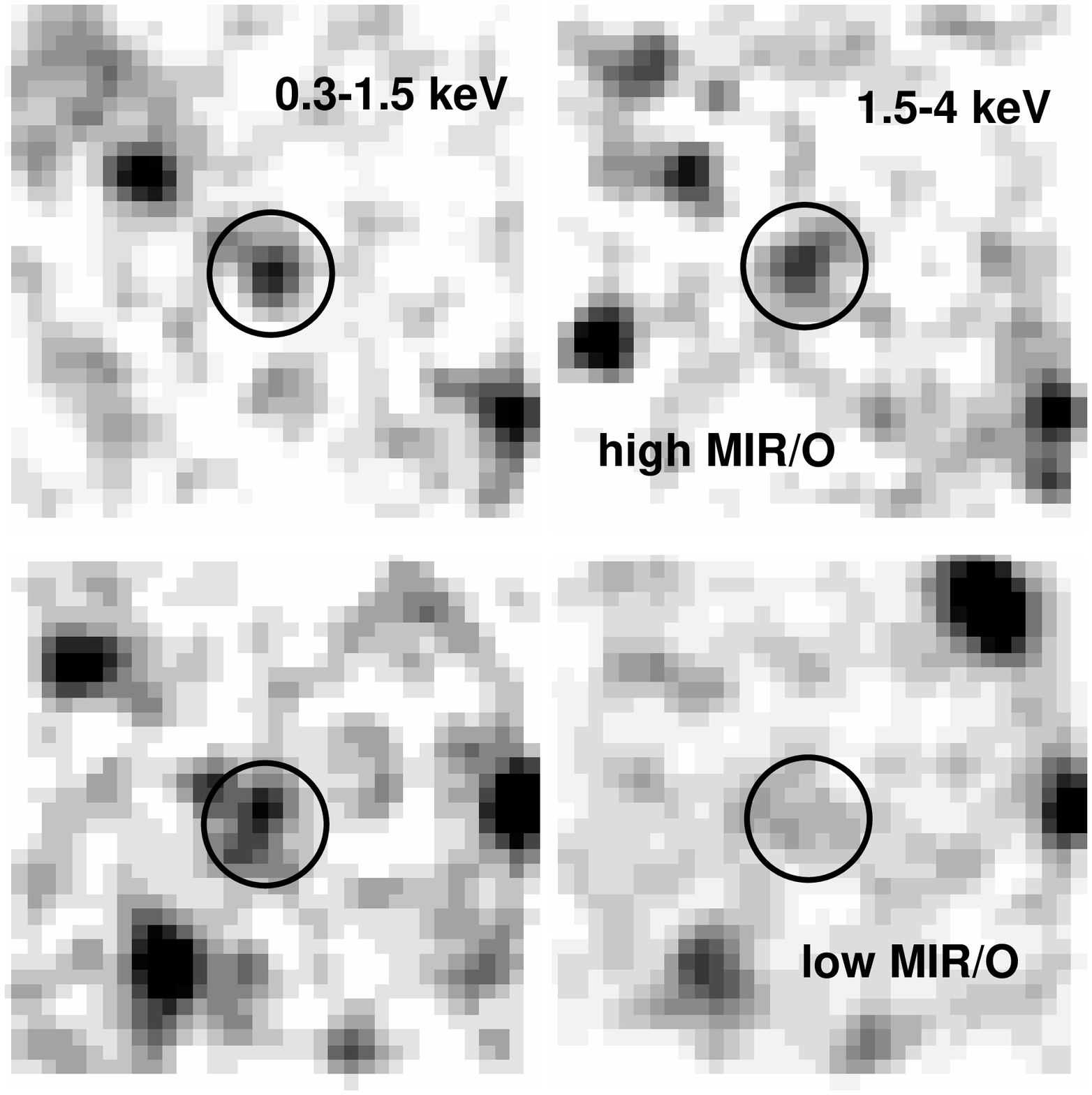}
\includegraphics[height=8.5truecm,width=8.5truecm]{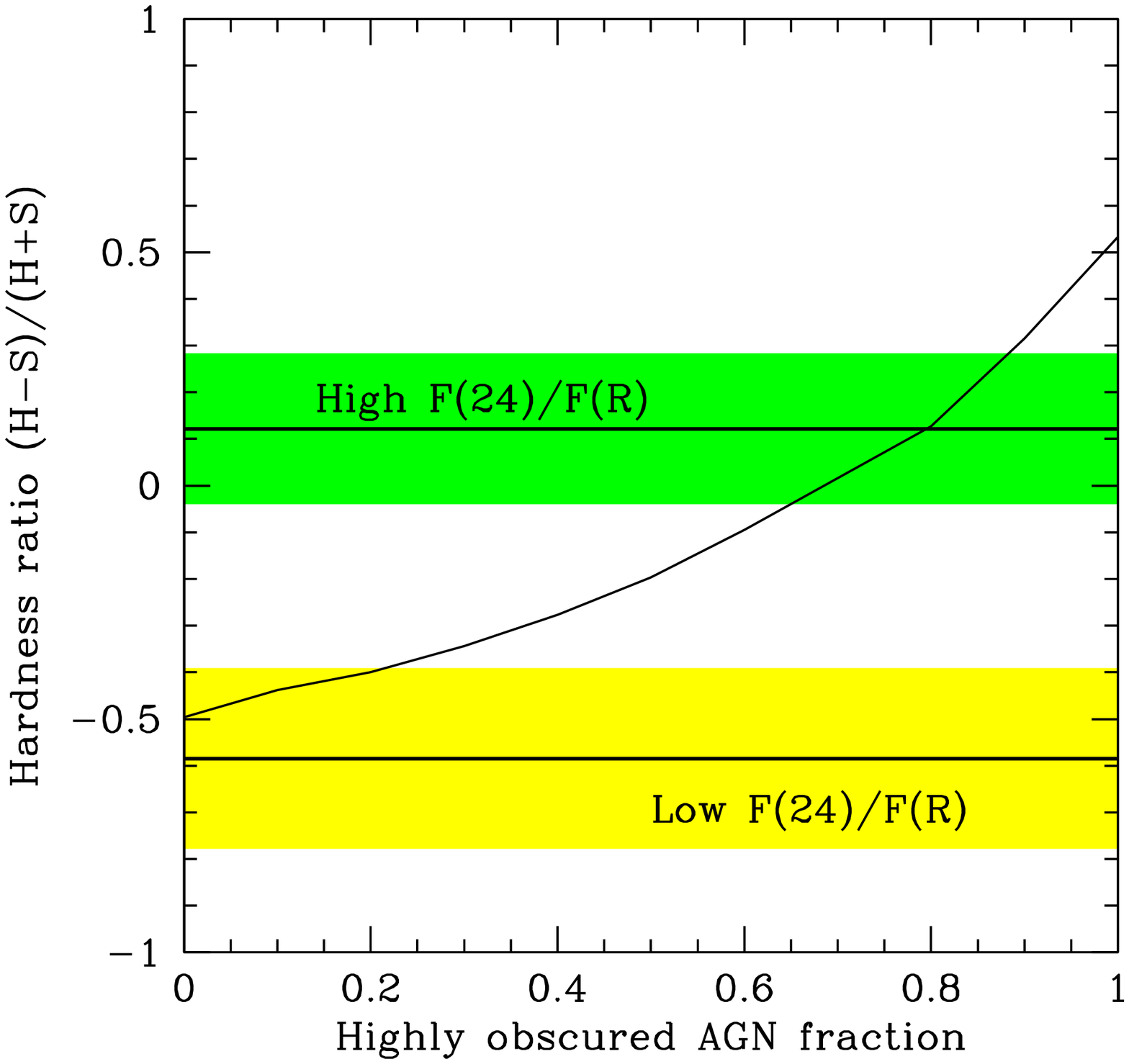}
\end{tabular}
\caption{Left figure: stacked X-ray images of 24$\mu$m GOODS-MUSIC
sources. Images have sides of 18 arcsec, central circles have 2
arcsec radii. Upper panels refer to 111 sources with
F(24$\mu$m)/F(R)$>1000$ and R-K$>$4.5; lower panels refer to 73
sources with F(24$\mu$m)/F(R)$<200$ and R-K$>$4.5. Left panels are
stacks in the S 0.3-1.5 keV band, right panels are stacks in the H
1.5-4 keV band. Right figure: the hardness ratio (H-S)/(H+S) of the
counts in circles of 2 arcsec radii as a function of the fraction of
highly obscured AGN for the sample of 111 24$\mu$m GOODS-MUSIC sources
with F(24$\mu$m)/F(R)$>1000$ and R-K$>$4.5. The solid curve is the
result of Montecarlo simulations (see text for details); the two thick
horizontal lines are the average hardness ratios measured for the
F(24$\mu$m)/F(R)$>1000$ and R-K$>$4.5 (upper line) and
F(24$\mu$m)/F(R)$<200$ and R-K$>$4.5 sources (lower line). The colored
bands mark the hardness ratio statistical uncertainties.}
\label{hrstack}
\end{figure*}

\subsection{Sources without a direct X-ray detection}

The total number of sources with F(24$\mu$m)/F(R)$>1000$ and R-K$>4.5$
and no direct X-ray detection is 111 (we excluded two
sources which happen to lie within 5 arcsec from an X-ray source). 

Four of these sources have a spectroscopic redshift, 99 have
photometric redshift in the GOODS-MUSIC catalog. For four sources we
could only compute a lower limit to the redshift. In conclusion, we
have redshifts or limits for 107 sources.  Both median redshift and
infrared luminosity are similar to those of the 22 sources with a
direct X-ray detection.

The redshift and infrared luminosities distributions of the
F(24$\mu$m)/F(R)$>1000$ and R-K$>4.5$ sources are compared in figure
\ref{zdist} to those of the full GOODS-MUSIC 24$\mu$m and X-ray
selected samples. The F(24$\mu$m)/F(R)$>1000$ and R-K$>4.5$ sources
have a redshift distribution with median redshift 1.91 (0.30).
Excluding the lower limits $<$z$>$=1.9 (0.28). This distribution is
shifted toward higher redshifts than both the full GOODS-MUSIC
24$\mu$m source sample and the X-ray selected AGN sample (figure
\ref{zdist} left panel). The right panel of figure \ref{zdist} shows
that moderately obscured, X-ray selected sources are concentrated
below z=1.5 and span a range of infrared luminosities log$(\lambda
L_\lambda (5.8\mu m))\sim42-45.3$ Their median 2-10 keV and infrared
luminosities are 43.16 (0.56) and 43.84 (0.57) respectively.  On the
other hand, the F(24$\mu$m)/F(R)$>1000$ and R-K$>4.5$ sources are
concentrated between z=1.2 and z=2.6 and have infrared luminosities
from log$(\lambda L_\lambda (5.8\mu m))\sim43.5$ to log$(\lambda
L_\lambda (5.8\mu m))\sim45.5$, with a median 44.34 (0.30),
luminosities similar to that of the X-ray selected AGNs at the same
redshift.

Table 1 gives the results of the template fits to the observed SEDs of
the 107 sources with a redshift.  In 65\% of the cases AGN templates
provide the best fit. The majority of these SEDs are best fitted by
one of the templates in figure \ref{seds}, the others are best fitted
by one of the AGN templates of Polletta et al. (2007).  Only one SED
is best fitted by the template of a type 1 AGN.  36 SEDs are best
fitted by star-forming galaxy templates (in 30 cases the best fit is
obtained using powerful star-forming galaxy templates, like those of
Arp220 and NGC6090). In conclusion, the results of the SED fitting
analysis confirm that the majority of the sources with
F(24$\mu$m)/F(R)$>1000$ and R-K$>4.5$ may be highly obscured
AGNs. However, we remark that the results of the template fitting
should not be always considered a quantitative determination of the
nature of each single source, since in many cases different templates
can produce fits with similar $\chi^2$ in the observed optical to
24$\mu$m band, producing a degeneracy difficult to account for.
Therefore, the results of the template fittings should only be taken
as a qualitative indication of the population properties of the
samples.

\subsection{X-ray stacking analysis}

To validate our highly obscured AGN selection sample and assess
quantitatively the fraction of bona-fide AGNs in the
F(24$\mu$m)/F(R)$>1000$ and R-K$>4.5$ sample, we performed a detailed
``stacking'' analysis of the X-ray data of the 24$\mu$m selected
sources. Indeed, thanks to the low Chandra background, it is possible
to increase the effective exposure time and derive average properties
of undetected objects, using these stacking techniques: counts at the
positions of known sources are co-added in order to probe X-ray
emission substantially below the single source sensitivity limit. 

We performed a stacking analysis of the 111 sources not directly
detected in the Chandra 1 Msec X-ray image. As control samples we used
the 22 sources with F(24$\mu$m)/F(R)$>1000$ and R-K$>4.5$ and an X-ray
counterpart, a sample of sources with F(24$\mu$m)/F(R)$>1000$ and
R-K$<3.5$ (51 sources after the exclusion of the sources directly
detected in the X-ray image) and a sample of sources with
F(24$\mu$m)/F(R)$<200$ and R-K$>$4.5 (73 sources after the exclusion
of the sources directly detected in the X-ray image). The total
exposure times for the four source samples are 94 Msec, 19.5Msec, 43
Msec and 61 Msec respectively.

The top panels of figure \ref{hrstack}a) show the X-ray stack of the
111 24$\mu$m sources with F(24$\mu$m)/F(R)$>1000$ and R-K$>4.5$ in two
energy bands. This is compared to the stack of 73 sources with
F(24$\mu$m)/F(R)$<200$ and R-K$>4.5$ and no direct X-ray detections
(lower panels).  We choose the bands 0.3-1.5 keV (soft band, S) and
1.5-4 keV (hard band, H) to keep the level of the internal background
as low as possible and similar in the two bands. At a typical
redshift of 2 these bands correspond to rest frame energies of 0.9-4.5
keV and 4.5-12 keV respectively.  The stack of the high
F(24$\mu$m)/F(R) and high R-K sources produces a detection in both
soft and hard bands, with count rates $(0.98\pm0.20)\times10^{-6}$
counts/s and $(1.25\pm0.26)\times10^{-6}$ counts/s in the two bands
respectively (using a 2 arcsec radius extraction region). The
hardness ratio H-S/H+S measured for this sample is therefore
0.12$\pm$0.15. Conversely, no detection in either bands is obtained
for the stack of 51 sources with high F(24$\mu$m)/F(R) and low R-K,
with 1$\sigma$ upper limits of $\sim2\times10^{-7}$ counts/s.  The
stack of the low F(24$\mu$m)/F(R) and high R-K sources produces a
significant detection only in the soft band, with count rate
$(1.89\pm0.32)\times10^{-6}$ counts/s. The count rate measured in the
hard band is significant at less than 3 $\sigma$ and corresponds to 
$(4.9\pm1.9)\times10^{-7}$ counts/s. The
corresponding hardness ratio is H-S/H+S=-0.58$\pm$0.18.  The stack of
the 22 sources with F(24$\mu$m)/F(R)$>1000$ and R-K$>4.5$ and an X-ray
counterpart produces a hardness ratio H-S/H+S$=-0.06\pm$0.01. In
conclusion, the stack of the 111 sources with F(24$\mu$m)/F(R)$>1000$
and R-K$>4.5$ and without a direct X-ray detection produces a
significant signal in both soft and hard X-ray bands. It is
interesting to note that its hardness ratio suggests an average
spectrum harder than even the average spectrum of the 22 sources with
similar IR and optical colors but with a direct X-ray detection.

\subsection{Simulations to assess the fraction of obscured AGN in the 
24$\mu$m source samples}

We used the observed flux in the stacked images, together with the
hardness ratio H-S/H+S, to constrain the fraction of highly obscured
AGN in the samples. To this purpose, we generated simulated X-ray
count rates and hardness ratios as a function of the fraction of the
AGN, assuming that the F(24$\mu$m)/F(R)$>1000$ and R-K$>4.5$ source
sample is made by obscured AGNs and star-forming galaxies.  We started
from the observed redshift and infrared luminosities, and for the
obscured AGN we assumed a log($\lambda L_\lambda
(5.8\mu$m)/L(2-10 keV)) luminosity ratio chosen randomly in the range
0.4-1.2. The lower value is the typical ratio found by Silva et
al. (2003) for low luminosity Seyfert 2 galaxies with column density
log$N_H\ls 24$. It is also similar to the ratio found for the two
HELLAS2XMM sources A2690\_75 and BPM16274\_69. The upper value is the
ratio found for the powerful obscured QSO IRAS09104+4109 (Piconcelli
et al. 2007).

For the star-forming galaxies we used a log($\lambda L_\lambda
(5.8\mu$m)/L(2-10 keV)) luminosity ratio between 2 and 2.8.  These two
values are obtained assuming a total infrared to 2-10 keV logarithmic
luminosity ratio of 3.6 (Ranalli et al. 2003) and correcting this for
the ratio between the total infrared luminosity and the 5.8$\mu$m
luminosity of powerful star-forming galaxies and spirals templates. 

We further assumed that the star-forming galaxies are not obscured in
X-rays, while the AGNs are highly obscured. For the latter
objects, we adopted a flat logN$_H$ distribution from 23 to 26
cm$^{-2}$.

For a given fraction of AGN in the sample, we first decide, using a
random generator, whether each source is an AGN or a star-forming
galaxy. Then, if the source turned out to be an AGN, we choose an
absorbing column density using again a random generator and the
assumed logN$_H$ distribution. We computed unobscured 2-10 keV
luminosities from infrared luminosities, and then X-ray fluxes by
folding a power law spectrum with spectral energy index of 0.8 reduced
at low energy by photoelectric absorption with a Chandra response
matrix. For column densities $\gs10^{25}$ cm$^{-2}$ we assumed that
the direct emission is completely blocked by photoelectric absorption
and Compton scattering. For these sources we assumed a reflection
component with normalization 1/100 of the direct component and same
spectral index.

A small fraction of the simulated count rates (20-30\%) is higher than
the Chandra detection limit. This was expected, given the fact
that 22 out of 135 sources with F(24$\mu$m)/F(R)$>1000$ and R-K$>4.5$
do have X-ray counterparts.  Since we are interested in the fraction
of AGN in the sources not directly detected in the Chandra images,
these count rates have been excluded from the following analysis.

The result of the simulations is shown in figure \ref{hrstack}, right
panel. According to our simulations, the average hardness ratio of the
F(24$\mu$m)/F(R)$>1000$ and R-K$>4.5$ source sample without X-ray
detection is reproduced if 80$\pm$15\% of the sources in the sample
are highly obscured AGNs. The hardness ratio of the sources with
F(24$\mu$m)/F(R)$<200$ and R-K$>4.5$ is reproduced if the fraction of
the obscured AGN is in the range 0-20\%.  Changing the assumptions
made to produce the simulations within reasonable ranges changes only
slightly this result.
%To account for these systematic uncertainties we
%increase by an additional 5\% the confidence intervals of the AGN
%fraction.  

The result of the simulations concerning the F(24$\mu$m)/F(R)$>1000$
and R-K$>4.5$ source sample is well consistent with the
indication coming from the 1-24$\mu$m SED fitting with galaxy and AGN
templates presented in the previous section. 

\section{Discussion and Conclusions}

We selected a source sample with extreme mid-infrared to optical flux
ratio (F(24$\mu$m)/F(R)$>$1000) and red optical colors (R-K$>$4.5) in
the GOODS CDFS area. These sources are among the most luminous sources
in the GOODS 24$\mu$m sample, because of the correlation of
F(24$\mu$m)/F(R) with the infrared luminosity (see figures
\ref{mirolir} and \ref{zdist}, right panel). The fraction of these
sources not directly detected in the Chandra images produces a
significant stacked signal in both Chandra soft and hard X-rays bands.
A detailed analysis based on Montecarlo simulation shows that the
stacked count rates and hardness ratios can be reproduced if this
source population is dominated by highly obscured (N$_{\rm H}>{\rm
a~few}\times 10^{23}$ cm$^{-2}$) AGNs.  

This conclusion is further confirmed by the following
consideration. If we assume that this source population is dominated
by star-forming galaxies, we can use the observed infrared luminosity
and the dust-corrected UV luminosities to derive two different
estimates of star-formation rates. On one hand total infrared
luminosities have been converted to a star-formation rate using the
following formula: $SFR=L_{IR} erg s^{-1}/2.24\times 10^{43}
M_{\odot}/yr$, (Kennicutt 1998). On the other hand the UV
star-formation rate is an output of the fits of the observed SEDs with
synthetic models.  These models have been obtained using the
Bruzual \& Charlot (2003) models, parametrizing the star-formation
histories with a variety of exponentially declining laws (with
timescales ranging from 0.1 to 15 Gyrs) and metallicities (from
$Z=0.02Z_{\odot}$ to $Z=2.5Z_{\odot}$, see Grazian et al. 2006 and
references therein for further detail). Dust extinction is added to
all star-forming models, with $0<E(B-V)<1.1$ and a Calzetti
attenuation curve. Given the very red colors of our objects in the
optical bands, they are typically fit with star-forming templates with
a large dust extinction ($E(B-V)>0.5$ ), resulting in a star-formation
rate much larger than what derived by a simple, dust free UV-to-SFR
conversion.

The median logarithmic star-formation rate from infrared luminosity is
2.31 (0.59) $M_{\odot}/$yr. The median logarithmic UV star-formation
rate is 0.83 (0.80)$M_{\odot}/$yr, a factor 30 lower. The large
mismatch between these two estimates strongly suggests that the
infrared luminosity of this source sample is not dominated by
star-formation but rather by accretion. As a comparison, the median
infrared and UV star-formation rates of the sources with
F(24$\mu$m)/F(R)$<200$ and R-K$>$4.5 in our control sample are 1.26
(0.38) and 1.10 (0.30), consistent with each other, suggesting that
the infrared luminosity of these sources is not dominated by nuclear
accretion (in agreement with their X-ray hardness ratio, see previous
section).

We can now compare the number of highly obscured AGNs selected at
24$\mu$m and not detected in X-rays, with the number of unobscured and
moderately obscured AGNs directly seen in the X-ray images. To reduce
the importance of complex selection effects we limit this analysis to
the sources in the redshift bin 1.2-2.6. At z=2.6 the 5.8$\mu$m
luminosity corresponding to the limit of 40$\mu$Jy at 24$\mu$m is
log($\lambda L_\lambda(5.8\mu$m))=44.3-44.2 for Seyfert
2,1.8 galaxy SEDs, and 44.4-44.5 for the four SEDs in figure
\ref{seds}. Assuming a log($\lambda L_\lambda (5.8\mu$m)/L(2-10 keV))
luminosity ratio in the range 0.4-1.2 implies unobscured 2-10 keV
luminosities in the range 43.3-43.9.  This suggests to limit the
comparison to the sources with logL(2-10 keV)$>43$ and 
log($\lambda L_\lambda(5.8\mu$m))$\gs$44.2.  

In the 1.2-2.6 redshift bin there are at least 46$\pm$10 sources with
F(24$\mu$m)/F(R)$>$1000, R-K$>4.5$, log($\lambda
L_\lambda(5.8\mu$m))$\gs$44.2 and no direct X-ray detection (80$\pm15\%$
of 57 sources), which are probably highly obscured AGNs, and 44 X-ray
selected AGNs, 7 of which with broad lines in their optical
spectra. The number of 24$\mu$m selected, presumably highly obscured
AGNs, missed by the CDFS X-ray survey is therefore similar to
the number of X-ray selected AGNs.

The median 5.8$\mu$m luminosity of the 57 F(24$\mu$m)/F(R)$>$1000
and R-K$>4.5$ sources without direct X-ray detection in the redshift
and luminosity bins defined above is log($\lambda L_\lambda
(5.8\mu$m))=44.48 (0.30), implying a median unobscured 2-10 keV
luminosity in the range 43.3-44.1, adopting our assumption on the
infrared to X-ray flux ratio. At z=2.6 these luminosities translate to
X-ray fluxes between a few times $10^{-16}$ and $10^{-14}$ \cgs,
meaning that these sources would have been easily detected if they
were not highly obscured.

We have compared our findings with the predictions of the La Franca et
al. (2005) and Gilli et al. (2007) models. In the redshift bin
1.2--2.6 and in the area covered by our 24$\mu$m sample the La Franca
et al. (2005) luminosity function predicts about 75 AGNs with
logL(2-10 keV)$>43$, 20 of which with a column density $\gs10^{24}$
cm$^{-2}$. The absorption distribution adopted by the Gilli et
al. (2007) AGN synthesis model for the Cosmic X-ray Background
predicts about 40 Compton Thick AGNs with the same limits for redshift
and luminosity.  We find 46$\pm$10 24$\mu$m selected, highly obscured
AGNs, a number about twice the La Franca et al. (2005) prediction but
well consistent with the Gilli et al. (2007) prediction.  

Marconi et. al. (2004, 2007 in preparation) derived a SMBH mass
function from the X-ray selected AGN luminosity functions that falls
short by a factor of about 2 to the ``relic'' SMBH mass function. This
difference can be greatly alleviated adding the population of infrared
selected, highly obscured AGNs to the AGN selected in the X-ray band
below 10 keV.

Finally, we note that the population of infrared selected, highly
obscured AGNs, can help in reconciling the predictions of models of
galaxy evolution with the observed AGN number densities at
z$\gs1-1.5$. Indeed, the Menci et al (2004) model predicts a number of
low-to-intermediate luminosity AGNs at z=1.5-2.5 about twice with
respect to that measured through X-ray 2-10 keV surveys (La Franca et
al. 2005, Fiore 2006). We found above that in the redshift bin 1.2-2.6
the number of X-ray selected AGN with logL(2-10keV)$>43$ is comparable
to that of the infrared selected, highly obscured AGNs. Therefore,
adding together the two populations would give a total number density
of AGNs similar to that predicted by the Menci et al. (2004)
model. More detailed, quantitative, AGN number density calculations,
spanning wider redshift and luminosity ranges, are beyond the purpose
of this paper and will be presented elsewhere (Fiore et al. in
preparation).

Most of the GOODS-MUSIC 24$\mu$m selected sources with high
F(24$\mu$m)/F(R), high R-K and no X-ray detection have a very faint
optical and near infrared counterpart, only 14\% of the sample having
R$\ls26$. Furthermore, only a handful of objects have a mid infrared
flux high enough (F(24$\mu$m$\gs0.5$mJy) to allow Spitzer IRS
spectroscopy.  The spectroscopic identification of $\sim$90\% of the
sample must therefore await the advent of ELTs or the launch of JWST.

Our analysis is limited to AGNs of intermediate luminosity at
z$\sim2$. To extend the coverage of the luminosity-redshift plane
requires a complementary observation strategy. Particularly useful to
this purpose is the SWIRE survey, which has both Spitzer and optical
medium-deep coverage on $\sim 50$ deg$^2$ of sky.  As an example, the
SWIRE survey contains hundreds of sources with extreme
F(24$\mu$m)/F(R) flux ratios and red optical-near infrared colors.
Interestingly, several dozens of these extreme sources have 24$\mu$m
flux higher than $\sim 1$ mJy, allowing Spitzer IRS spectroscopy,
and/or optical magnitude brighter than R$\sim25$, allowing
spectroscopy with 8m class telescopes. Furthermore, all these sources
are well within the reach of Herschel instruments between 75 and 500
$\mu$m.  Such long wavelengths observations can greatly help in
separating nuclear activity from star formation when the two
components have comparable integrated luminosities, e.g. in low
luminosity, Seyfert like, AGNs.

\acknowledgments

We are grateful to Fabio La Franca and Roberto Maiolino for useful
discussions. We thank an anonymous referee for comments that helped
improving the presentation. After submission of this paper to ApJ we
became aware of a work by a different group reaching similar
conclusions (Daddi et al. 2007).  We thank Emanuele Daddi for
providing a copy of his manuscript before submission and for useful
discussions.  Part of this work was supported by ASI/INAF contracts
I/023/05/0 and I/024/05/0 and by PRIN/MUR grant 2006-02-5203.  Part of
this work was supported by the Deutsches Zentrum f\"ur Luft-- und
Raumfahrt, DLR project numbers 50 OR 0207 and 50 OR 0405.

\end{document}